\documentclass[a4paper,11pt]{article}
\usepackage{pos}
\usepackage{xspace}
\usepackage{sidecap}

\usepackage[backend=bibtex,maxcitenames=1,style=science]{biblatex}
\title{Towards the second H.E.S.S. Galactic plane catalogue}
 \ShortTitle{Towards the 2HGPS catalogue}

\author*[a]{Q. Remy}

\affiliation[a]{Max-Planck-Institut fur Kernphysik,\\
  Saupfercheckweg 1, 69117 Heidelberg, Germany}

\onbehalf{for the H.E.S.S. collaboration} 

\emailAdd{ quentin.remy@mpi-hd.mpg.de}

\abstract{
The H.E.S.S. Galactic Plane Survey (HGPS), carried out between 2004 and 2013, is the most extensive survey of our Galaxy at very-high energies that covers the southern sky. Since the first HPGS catalogue release, the new observations accumulated provide a deeper scan of many Galactic sources, and a number of improvements have been made at various stages of the data processing chain, notably on events reconstruction and background modeling techniques. In parallel a new catalog production workflow has been tested and optimized on simulations done in preparation for the future Galactic Plane survey to be conducted by the Cherenkov Telescope Array (CTA). The development of a common data format and open scientific tools for gamma-ray astronomy allowed a smooth transition from the exploratory work done on CTA simulations to its application for H.E.S.S. data analysis. These elements offered a solid ground to build the second H.E.S.S. Galactic Plane Survey catalogue (2HGPS). In the following we will focus on the description of the catalogue workflow and show few results along the way.
}

\ConferenceLogo{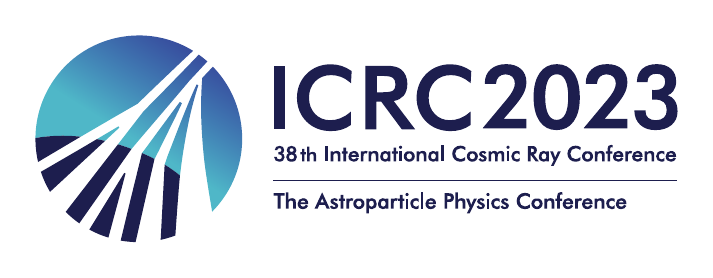}

\FullConference{%
38th International Cosmic Ray Conference (ICRC2023)\\
  26 July - 3 August, 2023\\
  Nagoya, Japan}

\newcommand{\citep}{\cite}
\newcommand{\citet}{\cite}

\newcommand{\hess}{H.E.S.S.}

\newcommand\change{}

\newcommand\bkg{{\change CR background}\xspace}

\newcommand{\BOout}[1]{}

\newcommand{\MCout}[1]{}

\begin{document}
\maketitle

\section{Introduction}

For the 2HGPS catalogue we aim to provide a consistent description of the whole Galactic plane observable by H.E.S.S., through a fully reproducible procedure. Compared to the first HGPS \cite{2018A&A...612A...1H} it features a more sophisticated modelling of the Cosmic rays (CR) background, diffuse emission, and sources.
The catalogue workflow was build using \textit{gammapy} \cite{gammapy17}, and directly inherited from the work done in preparation of the CTA-GPS survey \citep{2022icrc.confE.886R}. The tests performed on the CTA-GPS simulation shown that we can produce a catalogue close to the absolute limit of detections expected from the true sky models. For the application to H.E.S.S data several improvements were added regarding the modelling of the backgrounds (Sect.\ref{sec:bkg_norm}), and the selection of the sources in the final catalogue (Sect. \ref{sec:cat_opti} and \ref{sec:final}). The structure of this article follows the main steps of the catalogue workflow.

\section{Data reduction}

\subsection{Observations and events selection}

\begin{figure*}
    \centering
    \includegraphics[width=\textwidth]{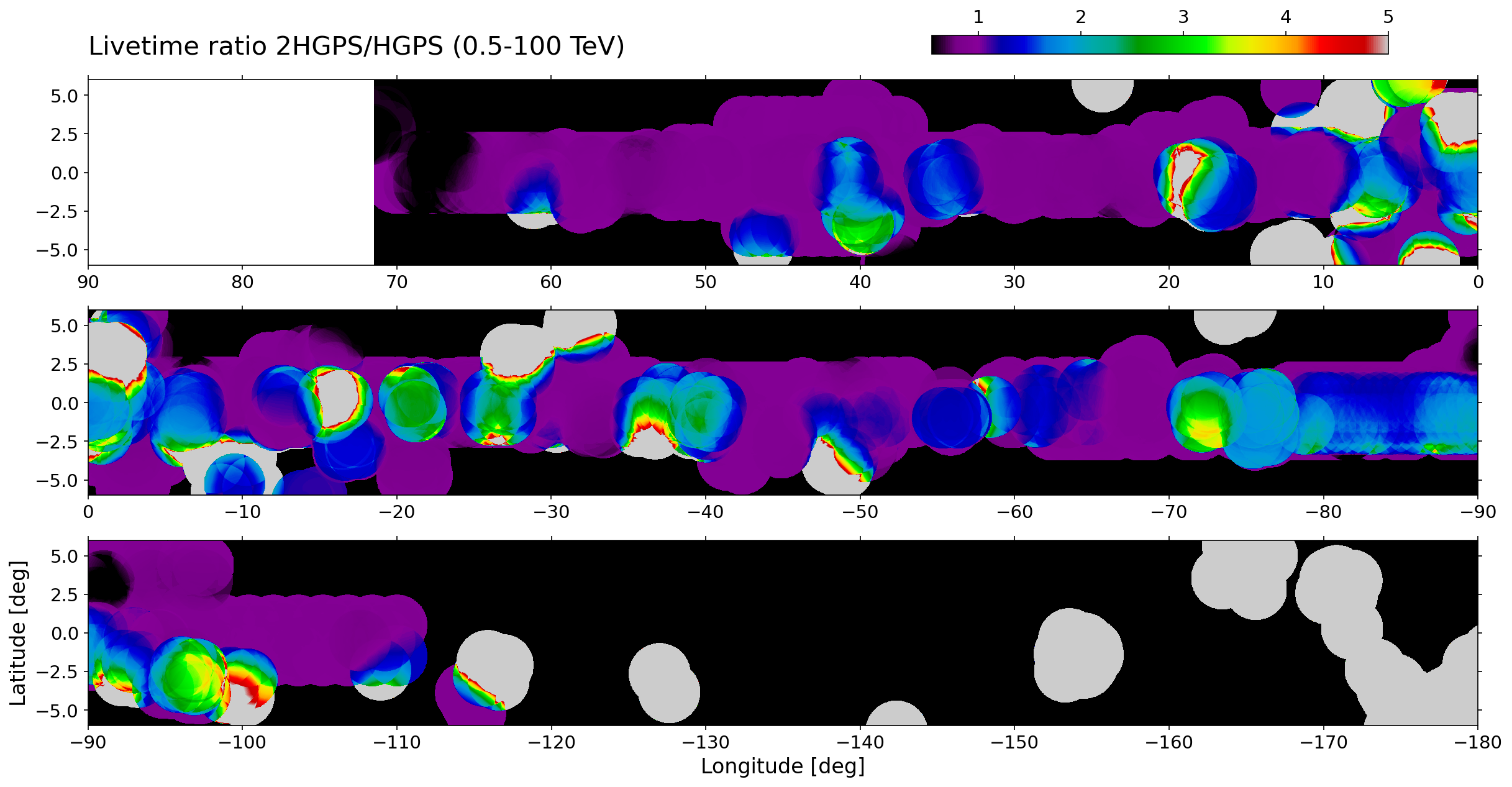}
    \caption{Livetime ratio between the observations used for the 2HGPS and the HGPS catalogues.}
    \label{fig:livetime}
\end{figure*}

This analysis starts with events list and instrument response functions from ImPACT reconstruction  \cite{2014APh....56...26P} in GADF-DL3  format \cite{2021Univ....7..374N}.
Observations in the Galactic Plane within -190$^\circ$ < l <72$^\circ$ and $|b|<6^\circ$ are included if they have a zenith angle lower than 65$^\circ$, a data quality flag "zero", and a field-of-view background model available \cite{2019A&A...632A..72M} (up to 2019).
The events are binned into maps with a spatial pixel width of 0.03$^\circ$ and 10 bin per decade logarithmically spaced in energy from 500 GeV to 100 TeV. Only events within a 2$^\circ$ offset from the observation pointing are selected. Energies where the effective area is less than 10 percent of its maximum are also masked.
As shown in Fig. \ref{fig:livetime}, the new data addition compared to the first HGPS catalogue offers deeper observations of many sources, several wide scan of key regions like the Galactic center and Carina arm, and observations in the outer Galaxy.

\subsection{Background normalisations scaling}\label{sec:bkg_norm}

\begin{figure*}
    \centering
    \includegraphics[width=\textwidth]{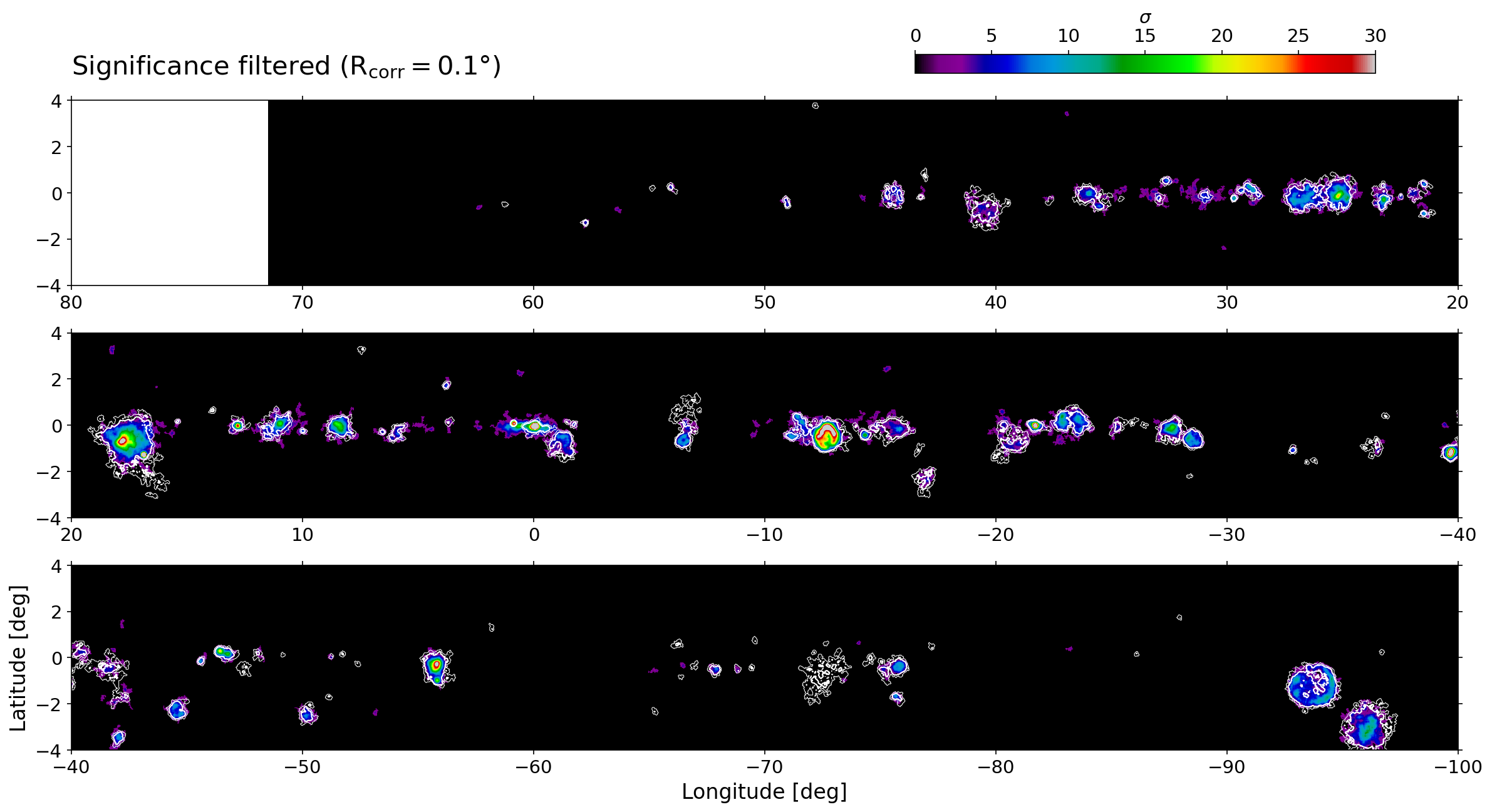}
    \caption{Filtered significance of the excess above the \bkg. The map correspond to the result of the HGPS (integrated above 1 TeV) and the contours show the 2HGPS significance at 3 and 8$\sigma$ (integrated in the 0.5-100 TeV energy range). The spatial bins width is $0.02^\circ$ for both.}
    \label{fig:signi}
\end{figure*}

The normalisation of the CR background model is re-fitted in each energy bin for each observation outside an exclusion mask. Initially the exclusion mask includes: circular regions of 0.3$^\circ$ radius around each known source listed in gamma-cat\footnote{\url{https://gamma-cat.readthedocs.io/}}, a $|b|<0.5^\circ$ band, and a mask derived from the significance of the excess above the background. For the latter the significance maps are filtered by hysteresis thresholding using the implementation provided by scikit-image \citep{scikit-image}. Two thresholds are used: first, pixels above the higher threshold are selected, and then pixels between the two thresholds are selected only if they are continuously connected to a pixel above the high threshold. The low and high thresholds are set to  2$\sigma$ and 4$\sigma$, respectively. The significance maps are computed in the $E=0.5-100$~TeV energy range for two correlation radii, R$_{corr}$ = 0.1$^\circ$, 0.2$^\circ$. The masks of selected pixels from these two maps are combined with an \textit{or} condition in order to update the exclusion mask. Note that in the first iteration we use the public significance maps from the HGPS paper, but once the datasets are produced we perform a second iteration using the new significance maps (see Fig. \ref{fig:signi} for a comparison).

The CR background normalisation will be biased upward if the diffuse Galactic emission is not taken into account as it extends beyond the exclusion mask, across the whole field-of-view. So we introduce an interstellar emission model (IEM) computed with HERMES code \cite{Dundovic:2021ryb}. We use the same CR transport setup as in the "varmin" model from \citet{Luque:2022buq}, but with a normalisation tuned such that the radial gradients across the Galaxy in both $X_{\rm CO}$ and gamma-ray emissivity per gas nucleon at 8 GeV match \textit{Fermi}-LAT measurements \citep{Acero2016apjs}.
As the IEM has been tuned to the \textit{Fermi}-LAT measurements but not to the H.E.S.S data, it has to be refitted. As a minimal preliminary step, we refit the normalisation of IEM and CR background globally, together with the normalisation of the sources from the HGPS catalogue. The fit is performed a second time after masking the excesses not accounted by the HGPS sources (the mask is given by the hysteresis-filtered significance maps).

The pixels where the ratio of IEM to CR background counts is greater than 5\% are added to the exclusion mask instead of the $|b|<0.5^\circ$ band along the Galactic plane.
Finally, we repeat the fit of the CR background normalisation per observation with the updated exclusion mask and including the IEM in the models.

\section{Candidate objects}\label{sec:detect}

\subsection{Object detection}

After computing the hysteresis-filtered significance maps of the excesses above CR background and interstellar emission model,
the object detection is performed combining three methods:
\begin{itemize}
\setlength\itemsep{-0.4em}
\item Peak detection: find local maxima above 5$\sigma$;
\item Circle detection: the contours of each group of pixels isolated by the filtering are fitted as a circle. If less than \mbox{80 $\%$} of the pixels are included in the circle, the object is discarded;
\item Edge detection and Hough circle detection: details are given in \cite{2020APh...12202462R}. This method is applied only on the map with R$_{corr}=0.1^\circ$ for simplicity.
\end{itemize}

Removal of likely duplicate objects is performed when the results of the different detection methods are combined. Groups of objects with an inter-center angular separation less than  $0.1^\circ$ and a difference in radius less than $0.25^\circ$ are replaced by a single object by averaging their position and radii.

\subsection{Associations}\label{sec:asso}

In order to match the detected objects in the catalogue with the sources in previous catalogues we test for spatial coincidence using two criteria.
For each object we first select true sources within a given angular separation to the object centre defined as \mbox{$d_{\rm inter-center}<d_{min}+f_R \times R_{\rm object}$}, where  $R_{\rm object}$ is the 68\% containment radius of the fitted model convolved by the PSF.
We set $d_{min}=0.1^{\circ}$ and $f_R = 1$ which is equivalent to the association criterion used in te first HGPS catalogue.
After this first selection we report the association maximising the surface overlap fraction, defined as:
\mbox{$\rm{SF_{\rm overlap}}= \left( S_{\rm object\,  \cap  \, source} \right) / \left( S_{\rm object \, \cup \, \rm source} \right)$},
where the surfaces are delimited by the iso-contours at 68\% containment in flux of the model convolved by the PSF.

We choose to report only the association that maximises the surface overlap fraction in order to limit the possible associations for extended objects. We also enforce that each detected object can be associated with only one source per catalogue and vice-versa. Moreover, only the associations with \mbox{$\rm SF_{ overlap}>0.25$} are reported in order to limit spurious associations. We checked on a sample of mock catalogues that this threshold maximizes the balanced accuracy defined as the average of true positive and true negative rates for associations.


\subsection{Candidates selection}

Extended source morphologies are usually more complex than the parametric models we use to describe them, so a single extended source can be detected as a cluster of objects. To address this, we start by classifying the objects depending on their degree of overlap. The surface considered to estimated the overlap is as disk using the radius estimated from the detection step. In the following non-overlapping objects are referred as isolated, while parents are large objects partially overlapping with smaller ones. These smaller objects are considered as sub-structures of their parent. 

For each object, a baseline spatial model is determined using the Pearson correlation coefficient ($PCC$) of a 5-point radial profile in flux. To this purpose, we integrated the flux map (with R$_{corr}$ = 0.1$^\circ$) in 5 rings of equal area between the centre of the object and its radius. Then we computed the PCC  between the radius and flux values. The default spatial model is a generalised Gaussian (see next section). Alternatively, a shell is considered first if the candidate object has $PCC\, >1/3$, or if it has $|PCC|\, <1/3$ and overlaps with objects classified as sub-structures.

For each object we compute a test statistic (TS) defined as the squared significance of the residual excess integrated within a correlation radius equivalent to the object radius. The candidate objects are filtered by requiring $ \rm TS>10$.
We then perform outlier detection using the isolation forest algorithm \citep{2018arXiv181102141H} implemented in scikit-learn \citep{scikit-learn}. The object properties considered when applying the outlier detection are: the mean distance of the 5 nearest neighbours, the distance of a sub-structure to its parent-object, and the PCC. This parameter space informs us on the expected density of objects with a given morphology. This information is used to train the outlier detection classifier and to set its selection score.
Sub-structures below a threshold in selection score are discarded in order to reduce spurious detections in complex sources (those that still remain after the removal of likely duplicate objects). The remaining objects are then ranked according to their selection score, which is used to determine their fitting order in the following (see next section).

Furthermore, after the TS filtering and the outlier detection, the selected candidates are divided into two lists: [primary] objects associated with known sources, isolated objects, parent objects, and sub-structures more significant than their parent with a difference in TS larger than 25; [secondary] unassociated sub-structures less significant than their parent.
These list provides source candidates, with robust guesses on their position and morphological parameters, to be tested subsequently by a conventional template-fitting analysis.

\section{Models optimization and selections}\label{sec:modeling}

\subsection{Default model fitting}

For model fitting, the Galactic plane is divided into 10$^\circ$ wide regions separated by 5$^\circ$ (half-overlapping). A 3$^\circ$ margin is added to each sub-region to account correctly for the sources outside the analysis region that contribute inside due to the PSF.
Regions containing less than five sources are merged with their neighbour in order to limit the number of regions fitted. The sub-regions obtained are then fitted independently.
The objects with centres outside of the fit region, but within the  3$^\circ$ margin, are merged into a unique background component.
So, for each energy range, we have three background {\change components: CR background}, IEM, and sources centred outside the fitting region.

By default, candidate sources are fitted with a log-parabola as spectral model and a generalised-Gaussian as spatial model\footnote{for details see: \url{https://docs.gammapy.org/1.0.1/user-guide/model-gallery/}}.
The minimum size fitted is $0.03^\circ$ (about half the PSF radius at 1 TeV).
Alternatively, a shell is fitted as the spatial model based on the morphological estimate from the initial detection step (see previous section).

For each region, the candidates in the primary list are fitted first while those in the secondary list are added only if there is still a significant residual excess after the fitting of the primary candidates ($\rm TS_{excess}>25$).
Once the initial candidate lists are exhausted, more sources are added iteratively (up to 5 per region) at the position of the largest peak above $5\sigma$ in the residual significance map.
Finally, for each object, we compute the test statistic for the null hypothesis (no source) and keep only those with $\rm TS_{null}>25$.
Once the fitting of all regions is complete, we assemble the final global model. Models that appear in multiple overlapping regions are taken from the region where they are located within $2.5^\circ$ from the centre of the region.

\subsection{Models refinement}

At this point, we define new regions centered on each source and we compare generalised-Gaussian, shell, or point-like hypotheses for spatial models, and log-parabola or power-law hypotheses for spectral models.
The optimal parameters of a model are given by the likelihood maximisation. The selection among non-nested models is performed by minimisation of the Akaike information criterion \citep[AIC, ][]{1974ITAC...19..716A}. The selection among nested models is done from a likelihood ratio test (TS) such as the more complex model is chosen only if preferred at more than 2$\sigma$. According to Wilk’s theorem the significance is computed assuming that the TS follows a $\chi^2$ distribution with a degree of freedom equal to the difference in the number of free parameters.

In order to simplify the global model we also search for groups of sources that could be replaced by an simpler model. We search for clusters of sources whose positions best match a linear or circular pattern using the RANSAC algorithm \citep{Fischler1981RandomSC}, implemented in scikit-image \citep{scikit-image}. The clusters of sources forming a linear pattern were replaced by an elliptical generalised Gaussian if the difference in AIC between multiple models and one elliptical model was greater than zero. Similarly, the clusters of sources forming a circular pattern were tested against a shell, but we also considered a shell plus a Gaussian (that could best model a composite system made of a PWN and and SNRs from the same progenitor).
We also looked for sources surrounded by negative or positive residuals above $1\sigma$ with an ellipticity larger than 0.5, for which we tested an elliptical generalised Gaussian.

\section{Catalogue threshold optimization}\label{sec:cat_opti}

\begin{figure*}
    \centering
    \includegraphics[width=\textwidth]{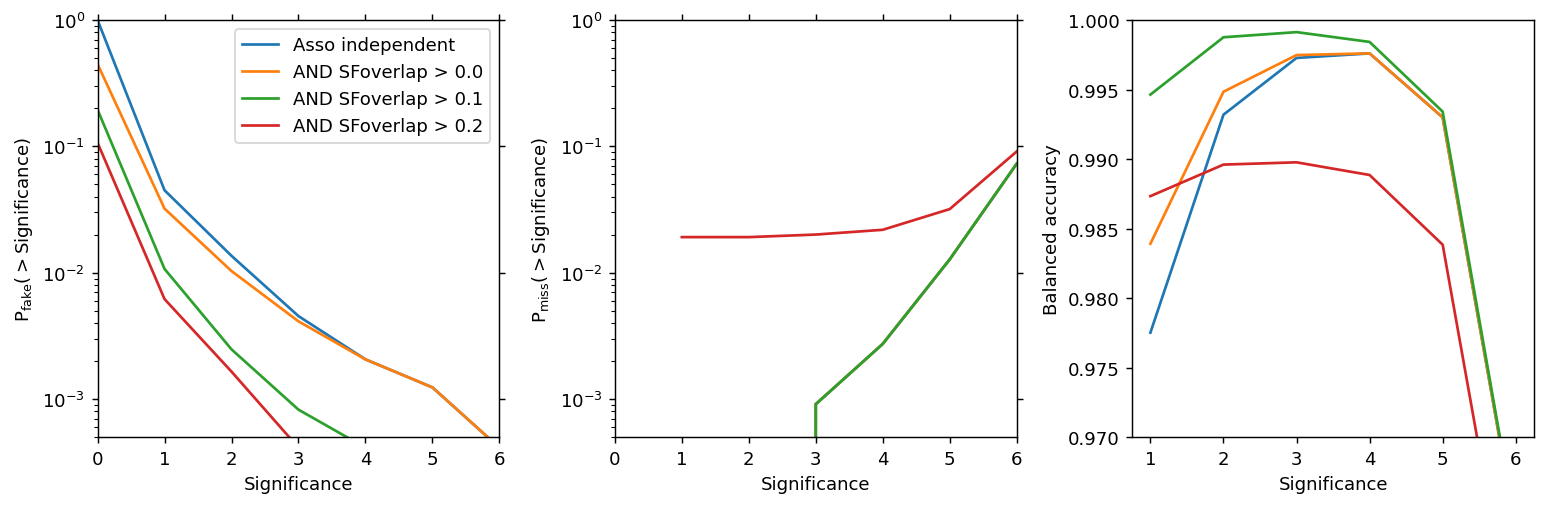}
    \caption{Probabilities of false positive, false negative and the balanced accuracy as a function of the Significance for different the association quality thresholds ($\rm SF_{overlap}$).}
    \label{fig:thr_opti}
\end{figure*}

The final catalogue threshold is estimated such as it minimizes fake and missed detection probabilities.
To do so we consider the detected objects associated to a source in the HGPS or gamma-cat catalogues as known. The remaining detected objects not associated to a known TeV source are used to generate lists of fake sources by boostrap, so they share the same properties (distribution in longitude, latitude, flux, size, and number). For each fake sky, the longitudes and latitudes of the objects in the original list are shuffled randomly and independently, excluding the positions that match the original ones.

Figure \ref{fig:thr_opti} shows the probabilities of false positive and false negative as a function of the Significance and the association quality ($\rm SF_{overlap}$), together with the balanced accuracy, defined as the mean of the true positive and true negative probabilities. The optimal threshold is given by the maximum in balanced accuracy. Based on this study we set the catalogue threshold to $4\sigma$ for candidates without association, and $3\sigma$ for candidates with association quality $\rm SF_{overlap}>0.1$. The objects that do not verify one of these conditions are removed from the models, after what the fitting and model selection operations are repeated. 

\section{Final catalogue}\label{sec:final}

Further tests are performed to flag unstable models considering different CR background parametrization, alternative diffuse models, and a cross-check dataset generated with a different events reconstruction technique.

Finally, we look for association within the final catalogue trying to match each object without association to know sources with one that is associated (using the procedure described in Sect. \ref{sec:asso}). These groups of objects are flagged as components of the same source.

\begin{SCfigure}
    \centering
    \includegraphics[width=0.76\textwidth]{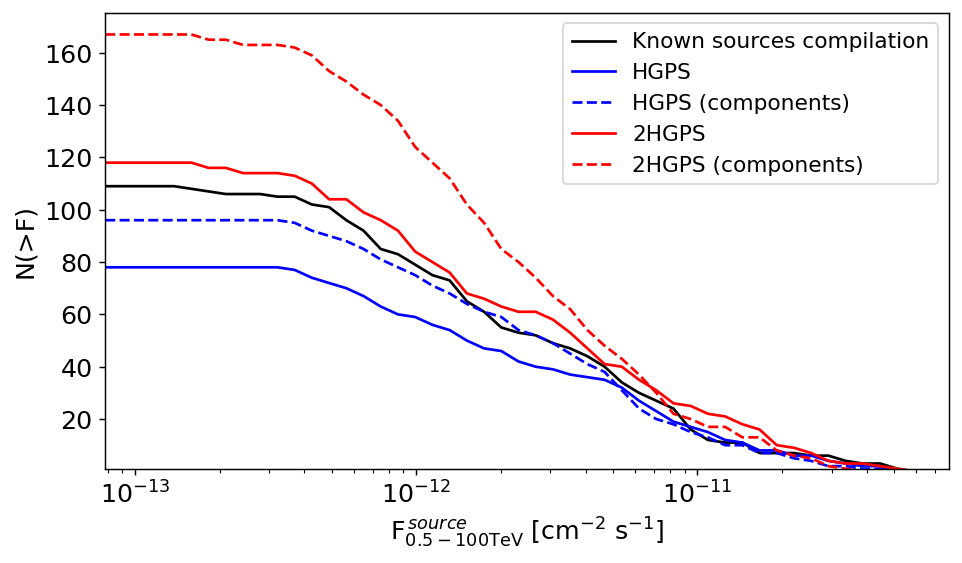}
    \caption{Number of objects above a given flux in the 2HGPS and HGPS catalogue compared to the compilation of known sources from \citep{2022icrc.confE.886R}. Note that the definition of sources (full lines) and components (dotted lines),  as well as the analysis regions, differ between the HGPS and 2HGPS catalogues.}
    \label{fig:NlogF}
\end{SCfigure}

\begin{SCfigure}
    \centering
    \includegraphics[width=0.75\textwidth]{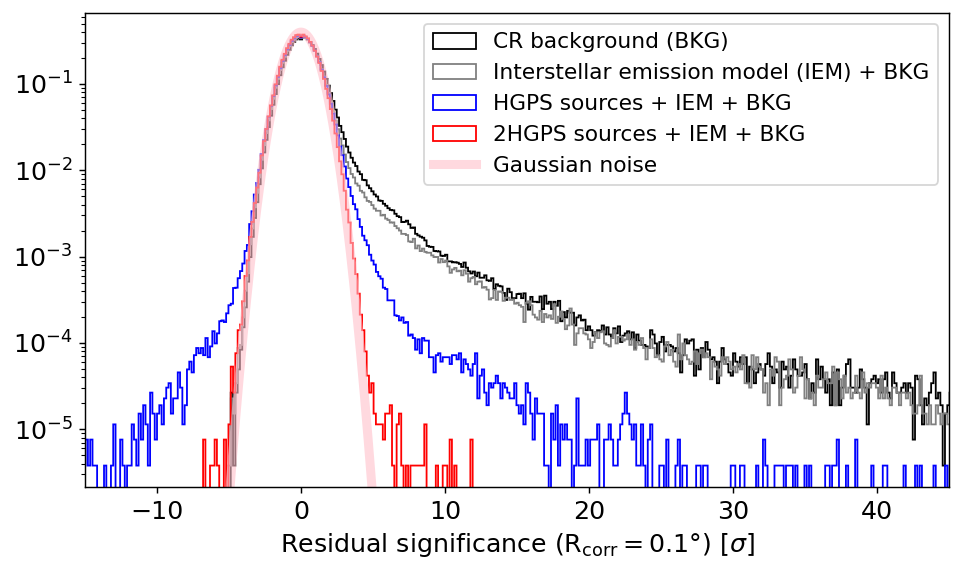}
    \caption{Distribution of the residual significance computed with R$_{corr}$ = 0.1$^\circ$ in the 0.5-100 TeV energy range considering different models.}
    \label{fig:resi}
\end{SCfigure}

The preliminary version of the 2HGPS catalogue contains about 120 sources modelled with 170 components. The distribution in number of objects above a given flux is shown in Fig. \ref{fig:NlogF}. As the 2HGPS catalogue provides several new detections, and a refined modelling of complex regions, the residuals excesses are clearly reduced compared to the HGPS (see Fig. \ref{fig:resi}). As a conclusion the upcoming 2HGPS catalogue will provide a new description of the $\gamma$-ray emission in the Galactic plane at very high energies with unmatched precision.

\scriptsize
\subsection*{Acknowledgments}
The \hess\ acknowledgments can be found at:
 \url{https://www.mpi-hd.mpg.de/hfm/HESS/pages/publications/auxiliary/HESS-Acknowledgements-2023.html}

\AtNextBibliography{\scriptsize}
\printbibliography


%
%
%

\subsection*{Full Authors List: \hess\ Collaboration}
\scriptsize
\noindent
F.~Aharonian$^{1,2,3}$, 
F.~Ait~Benkhali$^{4}$, 
A.~Alkan$^{5}$, 
J.~Aschersleben$^{6}$, 
H.~Ashkar$^{7}$, 
M.~Backes$^{8,9}$, 
A.~Baktash$^{10}$, 
V.~Barbosa~Martins$^{11}$, 
A.~Barnacka$^{12}$, 
J.~Barnard$^{13}$
% ...

\medskip

\noindent
$^{1}$Dublin Institute for Advanced Studies, 31 Fitzwilliam Place, Dublin 2, Ireland\\
$^{2}$Max-Planck-Institut f\"ur Kernphysik, P.O. Box 103980, D 69029 Heidelberg, Germany\\
$^{3}$Yerevan State University,  1 Alek Manukyan St, Yerevan 0025, Armenia\\
$^{4}$Landessternwarte, Universit\"at Heidelberg, K\"onigstuhl, D 69117 Heidelberg, Germany\\
$^{5}$IRFU, CEA, Universit\'e Paris-Saclay, F-91191 Gif-sur-Yvette, France\\
$^{6}$Kapteyn Astronomical Institute, University of Groningen, Landleven 12, 9747 AD Groningen, The Netherlands\\
$^{7}$Laboratoire Leprince-Ringuet, École Polytechnique, CNRS, Institut Polytechnique de Paris, F-91128 Palaiseau, France\\
$^{8}$University of Namibia, Department of Physics, Private Bag 13301, Windhoek 10005, Namibia\\
$^{9}$Centre for Space Research, North-West University, Potchefstroom 2520, South Africa\\
$^{10}$Universit\"at Hamburg, Institut f\"ur Experimentalphysik, Luruper Chaussee 149, D 22761 Hamburg, Germany\\
$^{11}$Deutsches Elektronen-Synchrotron DESY, Platanenallee 6, 15738 Zeuthen, Germany\\
$^{12}$Obserwatorium Astronomiczne, Uniwersytet Jagiello{\'n}ski, ul. Orla 171, 30-244 Krak{\'o}w, Poland\\
$^{13}$Department of Physics, University of the Free State,  PO Box 339, Bloemfontein 9300, South Africa\\
% ...

\end{document}